\documentclass{xsurv_article}
\usepackage{psfig}

\begin{document}

\begin{titlepage}
\setcounter{page}{1}
\makeheadline

\title {The X--ray background and the AGN X--ray luminosity function}

\author{{\sc G. Hasinger}, Potsdam, Germany \\
\medskip
{\small Astrophysikalisches Institut Potsdam} \\
\bigskip
}

\date{Received; accepted }
\maketitle

%%%%%\summary
%%%%% Text of summaryEND
\summary
ROSAT deep and shallow surveys have provided an almost complete inventory of
the constituents of the soft X--ray background which led to a population
synthesis model for the whole X--ray background with interesting
cosmological consequences. According to this model the X--ray background is the
``echo'' of mass accretion onto supermassive black holes, integrated over
cosmic time. A new determination of the soft X--ray luminosity function
of active galactic nuclei (AGN) is inconsistent with pure luminosity 
evolution.
The comoving volume density of AGN at redshift 2--3 approaches that of local
normal galaxies. This indicates that many
larger galaxies contain black holes and it is likely that the bulk of
the black holes was produced before most of the stars in the universe.
However, only X--ray surveys
in the harder energy bands, where the maximum of the energy density of the
X--ray background resides, will provide the acid test of this picture.
END
\keyw
surveys, active galaxies, cosmology
END

%%%%% Astronomy and Astrophysics Abstracts classification, type in if known
\AAAcla
END
\end{titlepage}

\kap{Introduction}

The soft X--ray background (XRB) has practically been resolved by deep ROSAT
pencil beam survey observations into discrete sources and -- at the faintest
fluxes -- source fluctuations (Hasinger et al.,
1993; Branduardi-Raymont et al.~1994). The ultradeep ROSAT HRI survey now 
reaches 
a surface density of $\sim 1000~deg^{-2}$ at a flux of 
$10^{-15}~erg~cm^{-2}~s^{-1}$ (Hasinger et al.~1998a). Counterparts of the 
weakest X--ray sources are optically very faint ($R<24$) and require very good,
unconfused X--ray positions
and high-quality optical spectra. The large majority of optical counterparts
 in the ROSAT deep
surveys turned out to be AGN (Schmidt et al.~1998).
X--ray selection is therefore the most efficient means to construct large,
almost unbiased samples of distant AGN. 

At the faintest X--ray fluxes there is still a debate about the existence of
a possible new population of nearby X--ray active, but optically innocent 
narrow emission
line galaxies (NELG), whose X--ray luminosity may be powered by star forming
processes (e.g. Griffiths et al.~1996; McHardy et al.~1997). 
These findings are, however, challenged by the higher resolution data in the 
Lockman Hole (Schmidt et al.~1998). 

\kap{The ROSAT Ultra-Deep Survey}

An area of $\sim 0.3 deg^2$ in
the Lockman Hole was chosen as the field for the deepest X--ray
survey ever, because it provides a minimum of
interstellar absorption. Over the last 10 years
it has become one of the best studied sky regions over
a wide range of frequencies: UBVRI and K-band imaging from
Kitt Peak, Mt. Palomar and Mauna Kea (Keck and UH), 
a 16hr VLA
20cm mosaique (deRuiter et al.~1997), a 120ksec ASCA
hard X--ray observation (Inoue et al.~1996) and
deep IR mosaiques with ISOCAM and ISOPHOT aboard ISO (PIs: Cezarsky,
Taniguchi) have been taken. The ROSAT Deep Survey (RDS) is a 207 ksec PSPC 
exposure, and the ROSAT Ultradeep Survey an 1.11 Msec ROSAT HRI exposure
in the Lockman Hole (Hasinger et al.~1998a).

\begin{figure*}[t]
\unitlength1cm
\begin{minipage}[t]{7.5cm}
\begin{picture}(7.5,7.5)
\psfig{figure=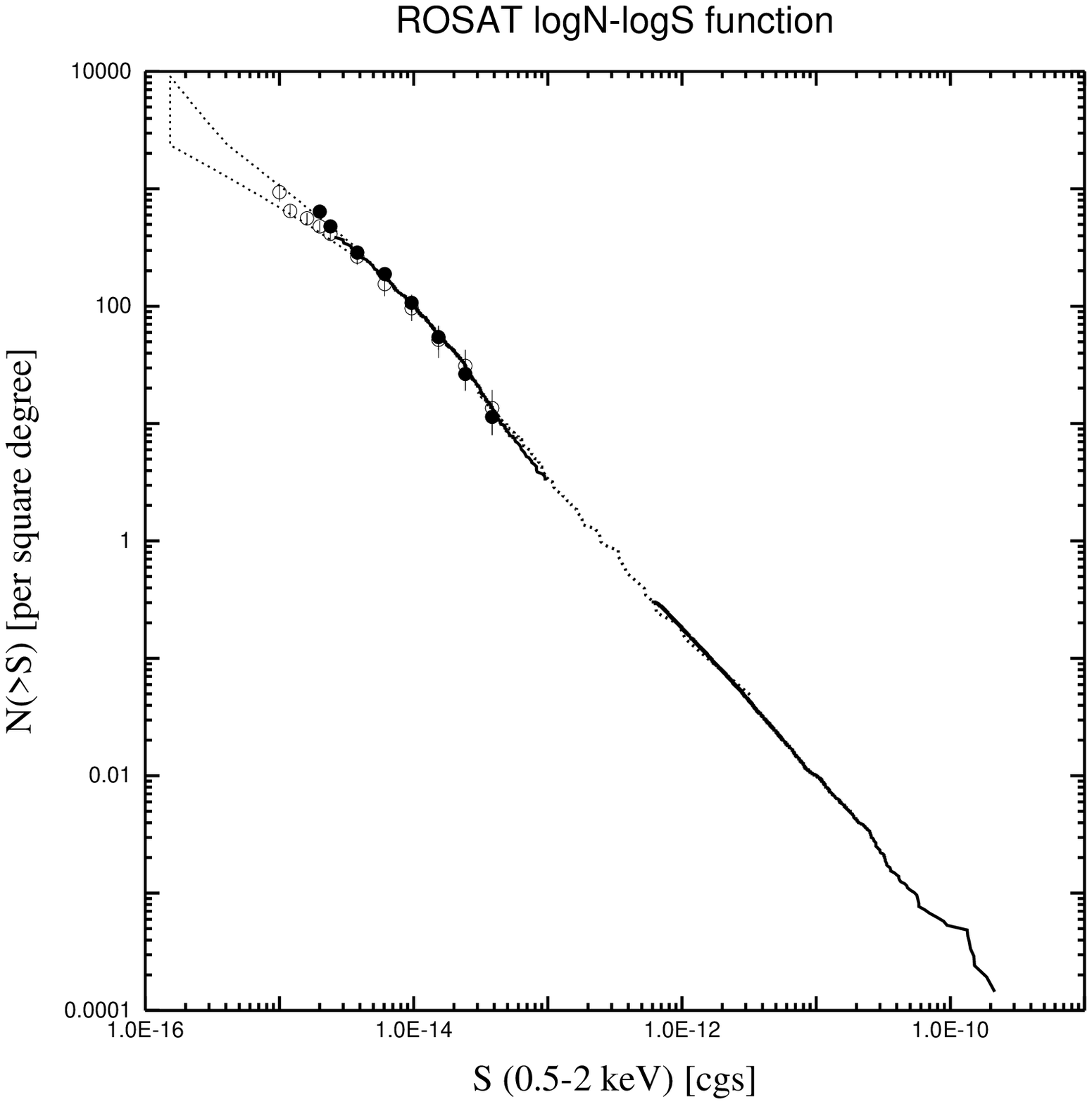,width=7.5cm} 
\end{picture}\par
\end{minipage}
\hfill
\begin{minipage}[t]{7.5cm}
\begin{picture}(7.5,7.5)
\psfig{figure=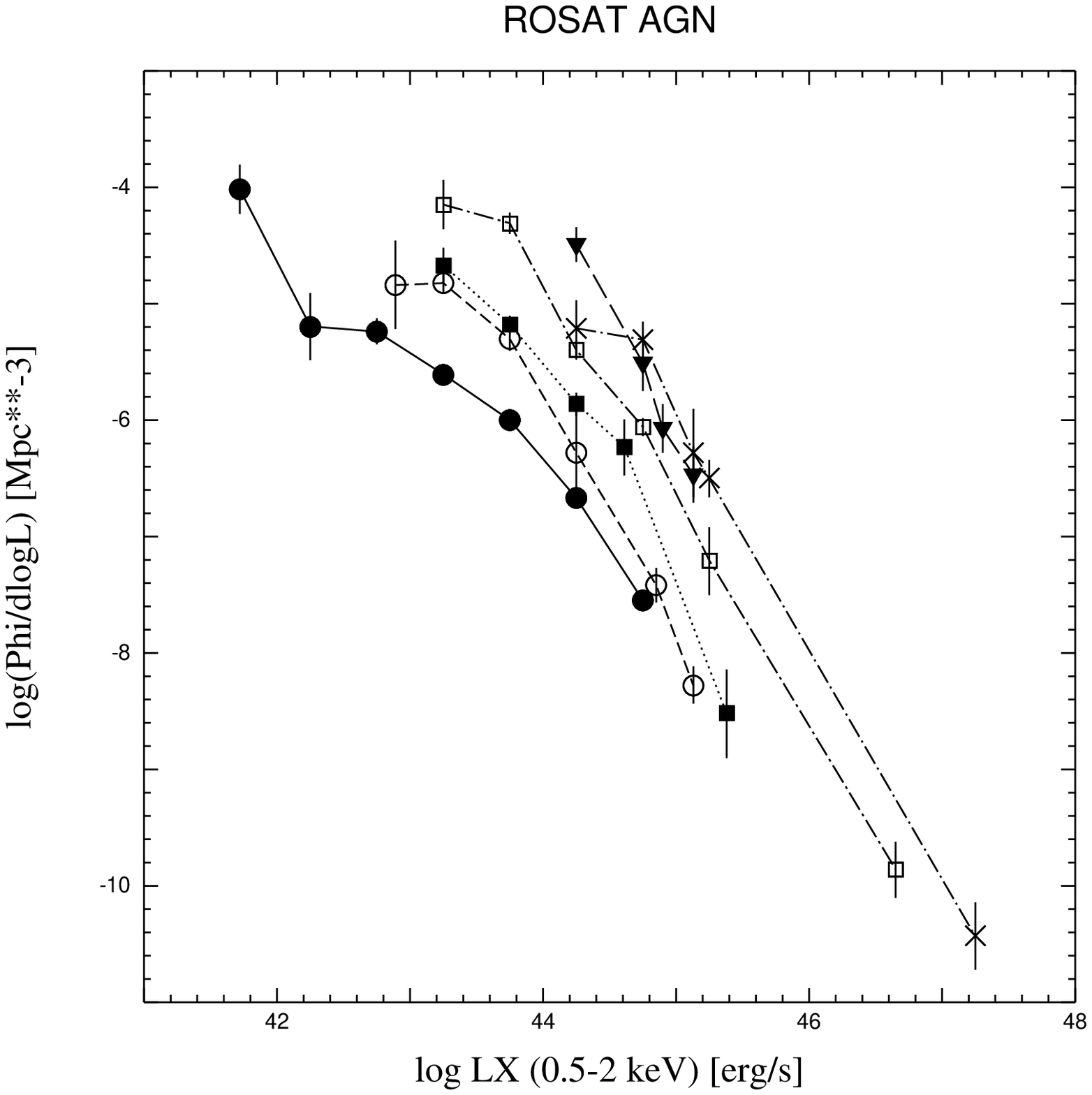,width=7.5cm}
\end{picture}\par
\vskip -0.2 truecm
\end{minipage}
\caption[]{\small 
a (left): 
Total log(N)--log(S) function determined from various
ROSAT surveys. Filled circles give the source counts from the 207 ksec PSPC 
observation in the Lockman Hole (Hasinger et al.~1998a). Open circles are 
from the HRI ultradeep survey observation (1.112 Msec). The data
are plotted on top of the source counts (solid line) and fluctuation
limits (dotted area) from Hasinger et al. (1993). The dotted line at
intermediate fluxes refers to the total source counts from the RIXOS
survey (Mason et al.~1996). The solid line at bright fluxes was determined
from the ROSAT All-Sky Survey bright source catalogue (Voges et al.~1995).

b (right):
The AGN X--ray luminosity function derived from a joint 
analysis of the ROSAT Deep and Ultradeep Survey, RIXOS and the ROSAT Bright 
Survey. The comoving AGN volume density is displayed as a bivariate 
function of luminosity and redshift. Redshift shells are 0-0.2 (filled
circles), 0.2-0.4 (open circles), 0.4-0.8 (filled squares), 0.8-1.6
(open squares), 1.6-2 (filled triangles), and 2-4.5 (asterisks).
}
\label{LNS}
\end{figure*}

The final limiting sensitivity for the detection of discrete sources
is about $2 \times 10^{-15}~erg~cm^{-2}~s^{-1}$
for the PSPC and about a factor of two fainter with the HRI.
Figure \ref{LNS}a shows the PSPC and HRI log(N)--log(S)
function (Hasinger et al.~1998a)
which is in very good agreement with data published previously. 
The HRI source counts reach a surface
density of $970\pm150~deg^{-2}$, about a factor of two higher than any
previous X--ray determination. About 70-80\% of the X--ray background
measured in the 1-2 keV band has been resolved into discrete sources.

\kap{Optical identifications}

Early optical identification programs concentrated on
medium-deep ROSAT PSPC observations in
AAT deep optical QSO fields and could quickly identify an
impressive fraction of faint X--ray sources as classical broad-line AGNs 
(mainly QSOs) (e.g. Shanks et al.~1991, Georgantopoulos et al.~1996). 
There was, however, mounting evidence that
a new class of sources might start to contribute to the XRB at faint
X--ray fluxes. The faintest X--ray sources in ROSAT deep surveys
on average show a harder spectrum than the identified QSOs
(Hasinger et al.~1993). In medium-deep pointings a number of optically 
``innocent'' narrow-emission line galaxies
(NELGs) at moderate redshifts (z$<$0.4) were identified as X--ray sources,
which was in excess
of those expected from spurious identifications with field
galaxies (Boyle et al.~1995, Georgantopoulos et al.~1995, Griffiths et al.,
1996). Roche et al. (1995) have
found a significant correlation of X--ray fluctuations with
optically faint galaxies. Finally, in an attempt to push optical
identifications to the so far faintest X--ray fluxes, McHardy et
al. (1997) claim that the surface density of broad-line AGN  
flattens dramatically
at fluxes below $5 \times 10^{-15}~erg~cm^{-2}~s^{-1}$,
while the NELG number counts still keep increasing, so that those would 
eventually dominate the XRB below a flux of $10^{-15}~erg~cm^{-2}~s^{-1}$.

While this is obviously an interesting possibility, it is useful to remind that
all these findings are based on identifications near the limit of
deep PSPC surveys, at fluxes where
our simulations suggest that the PSPC data start to be severely confused,
with some likelihood of misidentification (Hasinger et al.~1998a).
Also, spectroscopic optical identifications in these samples are limited to
$R < 22$.

In the Lockman Hole Deep survey,
optical counterparts have magnitudes in the range R=19--24.
With the excellent HRI positions (of order 2--4 arcsec) typically
only one counterpart is within the X--ray error box. Long--slit
and multislit spectra of the 1-3 candidates closest to the X--ray
source have been taken using the Palomar 200" 4-shooter 
and the Keck LRIS instruments. Reliable spectroscopic
identifications are now practically complete in 
a $0.30~deg^2$ field to a flux limit of $1.1 \cdot 10^{-14}~cgs$ and in 
the central $0.14~deg^2$ of the field down to a flux limit of
$5.5 \cdot 10^{-15}~cgs$ (Schmidt et al.~1998) and in a smaller area,
which we continue to work on, down to $10^{-15}~cgs$. The majority of the 
optical counterparts are active galactic nuclei with broad emission lines
(QSOs and Seyfert galaxies) in the redshift range 0.08-4.5, whereas
only a very small fraction of non-AGN NELGs is found. Data on the
highest redshift X--ray selected QSO will be presented elsewhere (Schneider et 
al.~1998). 
A large fraction of the faint X--ray sources are optically resolved
low luminosity AGN (Seyfert galaxies),
some of which show clear evidence of gas and dust obscuration in the X--ray and
optical band (i.e. Seyfert 1.5-2 galaxies), which are very hard to select by
any other means than X--rays.

The data of McHardy et al. (1997) is roughly consistent with our findings,
if we consider three selection/incompleteness effects, which are partly
discussed in their work: a.) their optical identifications are restricted
to $R<22$, while we have optical counterparts as faint as R=24. b.) they call
some fraction of objects NELGs, which due to our higher quality spectra
reveal some evidence for AGN activity, and c.) some fraction of their NELGs
may be misidentified due to source confusion.

\begin{figure*}[t]
\unitlength1cm
\begin{minipage}[t]{7.5cm}
\begin{picture}(7.5,7.5)
\psfig{figure=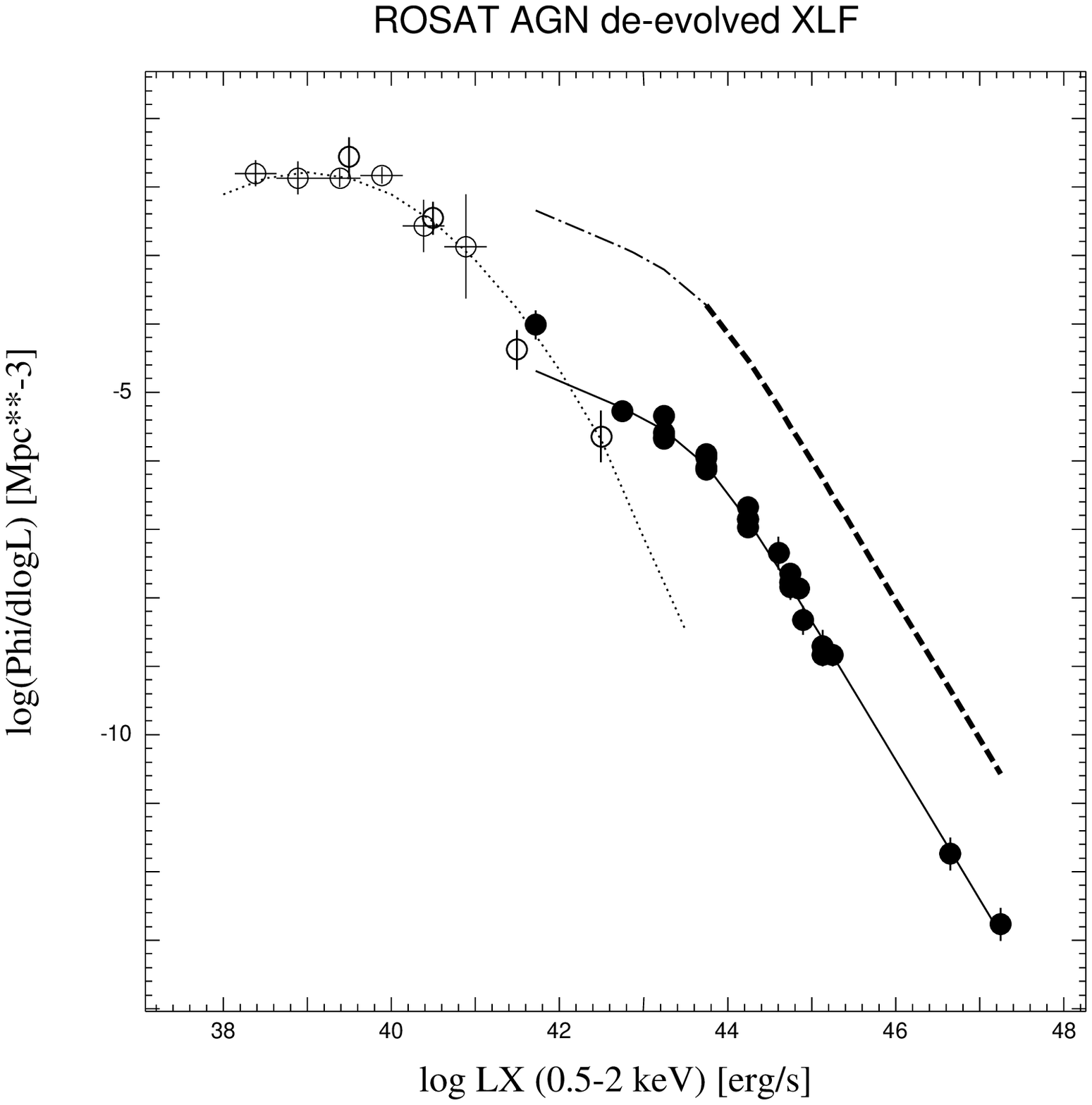,width=7.5cm}
\end{picture}\par
\end{minipage}
\hfill
\begin{minipage}[t]{7.5cm}
\begin{picture}(7.5,7.5)
\psfig{figure=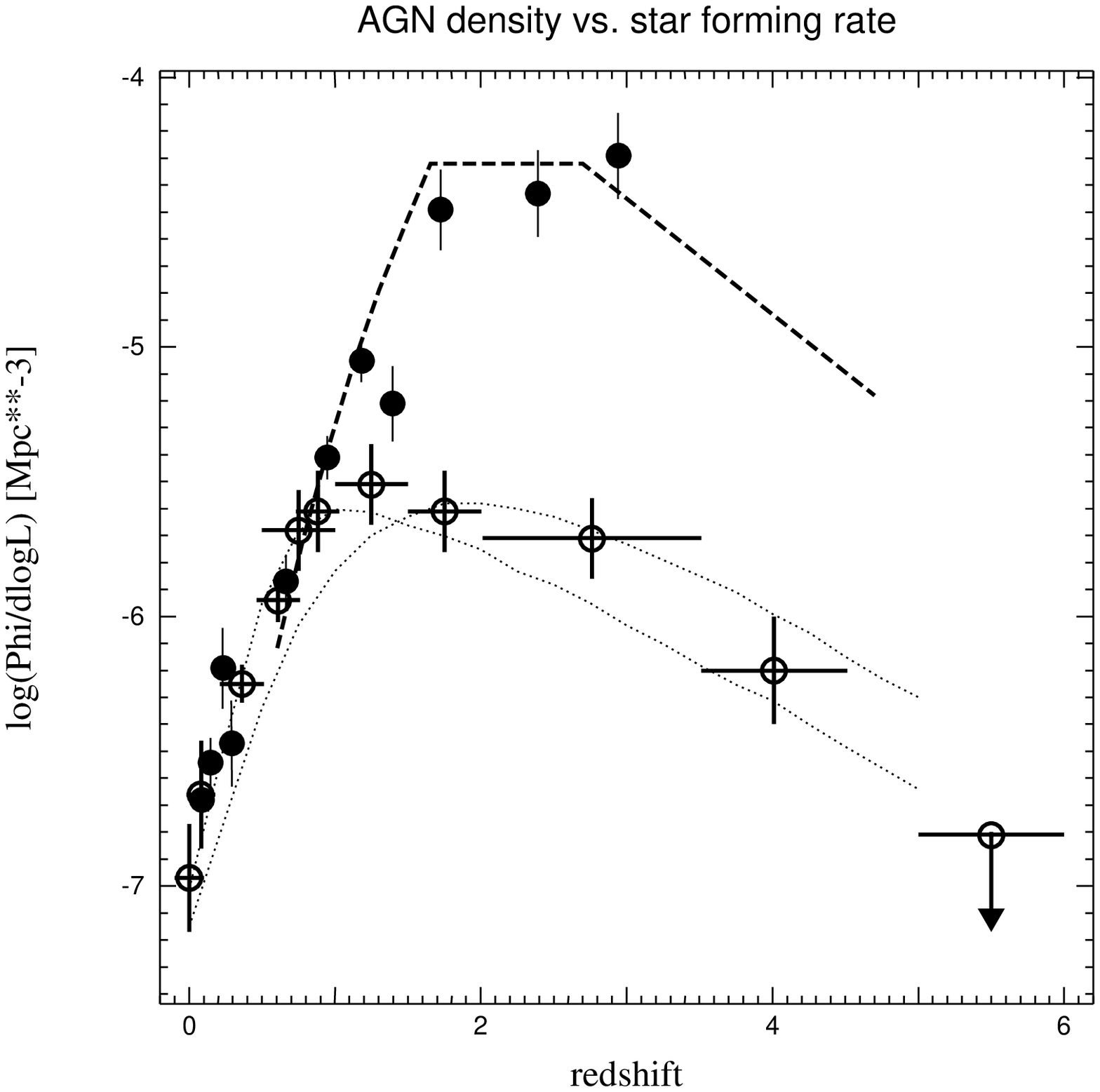,width=7.5cm}
\end{picture}\par
\vskip -0.2 truecm
\end{minipage}
\caption[]{\small 
a (left): 
X--ray Luminosity functions for AGN and galaxies from 
ROSAT surveys. The local galaxy luminosity function (open circles) has been
derived from the ROSAT Bright Survey (Hasinger et al.~1998b) and from
a volume-limited sample of local galaxies (Schmidt, Boller and Voges~1996).
A log-normal distribution has been fit to the galaxy XLF (dotted curve). The
AGN luminosity function (filled circles) has been de-evolved using a
pure density evolution model (see text). A smoothed broken power law model
(solid line) has been fit to the data. The dashed curve indicates the 
model AGN luminosity function observed in the redshift range 2-3, its
dash-dotted extension to lower luminosities is the model based on pure
density evolution (see text).

b (right): 
The filled circles give the volume density of X--ray selected 
AGN referred to
an X--ray luminosity of  $ log L_X = 44.25$. The dashed curve sketches the
volume density of optically selected QSO with $M_B < -26$ (Schmidt, Schneider
and Gunn~1995), increased by a factor of 250 to account for the different
luminosities in the different wavebands.
The open circles show the star formation rate as a function of redshift from
Madau 1996 and Connolly et al.~1997, corrected for dust obscuration
(Pettini et al.~1997). The dotted curves give models for the star formation
rate derived from the chemical evolution of $H_\alpha$ clouds (Pei and Fall
1995). The latter two curves are in arbitrary units, scaled to the AGN
density at low redshifts.
}
\label{MAD}
\end{figure*}

\kap{A new AGN X--ray luminosity function (XLF)}

Using data from medium deep ROSAT fields combined with the Einstein Medium 
Sensitivity 
Survey (Gioia et al.~1990), Boyle et al. (1994) could derive the AGN XLF and 
its cosmological evolution. Their data 
is consistent with pure luminosity evolution proportional to $(1+z)^{2.7}$ 
up to a redshift $z_{max} \approx 1.5$, similar to what was found previously 
in the optical range. This result has been confirmed and improved later on by 
more extensive or deeper studies of the AGN XLF, e.g. the RIXOS project
(Page et al.~1996) or the UK deep survey project (Jones et al.~1997).
All these studies agree that at most half of the faint X--ray source counts
can be explained
by classical broad-line AGN based on the pure luminosity evolution models.
However, the limited information at bright X--ray fluxes and in particular
the uncertain crosscalibration between the ROSAT and Einstein 
surveys severely limits the accuracy of the AGN XLF (Boyle et al.~1994;
Page et al.~1996).

We can now determine a new AGN soft X--ray luminosity function based on
ROSAT surveys alone and covering a very wide range of limiting fluxes. For
this purpose the deep survey sample described above has been combined with 
optical identifications from shallower wide angle
surveys, i.e. the ROSAT Bright Survey (RBS, Hasinger 
et al.~1998b), derived from the ROSAT All-Sky Survey bright source catalogue
(Voges et al.~1995), 
 and the RIXOS AGN sample (Page et al.~1996). The accuracy of the 
relative flux scale for these surveys is demonstrated by the logN--logS
function in Fig. \ref{LNS}a, where the different surveys agree within
$\sim 10\%$ in the overlap regions over a flux range of about six orders 
of magnitude.

Fig. \ref{LNS}b shows the binned AGN XLF in different redshift shells
(assuming $H_0=50~km/s/Mpc$, $q_0=0.5$).
Consistent with Boyle et al., we find strong cosmological evolution in the 
sense that high-redshift AGN 
are much more abundant or more luminous than their local counterparts. 
Contrary to the Boyle et al. findings, however, the new XLF is not
consistent with pure luminosity evolution. For the
first time we see evidence for strong cosmological evolution of the
space density of low-luminosity AGN (e.g. Seyfert galaxy) XLF out to a 
redshift 1-2, incompatible with pure luminosity evolution.
Surprisingly, however, a pure density
evolution model of the form $\Phi(z)=\Phi(0)\cdot(1+z)^{5}$ can fit the 
luminosity function well.

Fig. \ref{MAD}a shows the luminosity function, de-evolved under the
assumption of pure density evolution. A smoothed
broken power law model of the form
$$\Phi(L,z)/dlogL=(1+z)^{5.03} {{2\cdot 10^{-6}} \over {l^{2.03} + l^{0.51}}}
~[Mpc^{-3}]$$
has been fit to the de-evolved XLF out to redshifts of 1.92. Here $
l = {{L_X}/{5\cdot 10^{43}}}$. The model for z=0 is shown in figure \ref{MAD}a
by a solid line. The model for the z=2 XLF over the observed luminosity
range is shown by the thick dashed line, its extrapolation to lower 
luminosities by the dash-dotted line.  

Integrating the density evolution model of the luminosity function 
to $z\approx2$ actually overpredicts the soft 
X--ray background flux and the soft X--ray logN-logS function.
One has to stress however, that this is only a first-order treatment of the 
data.
The effect of intrinsic X--ray absorption and the detailed shape of the AGN 
spectra has to be included in the derivation of the XLF in order to obtain a 
self-consistent population synthesis model for the XRB spectrum (see 
e.g. Comastri
et al.~1995). Also, the so far unobserved low luminosity range 
of the high-redshift AGN XLF provides a significant contribution to the
XRB and the faint number counts and is therefore crucial in understanding 
the composition of the XRB. 
Even deeper observations, preferrably at higher X--ray energies will thus be 
necessary.

\kap{Does every galaxy contain a black hole?}

This question has originally been posed by Rees (1989). With recent
data from HST and from the deep X--ray surveys
we can now shed new light on this question. 
In fig. \ref{MAD}b we plot the volume density evolution of X--ray selected 
AGN (this work) and optically selected AGN as a function of redshift 
(Schmidt, Schneider \& Gunn 1995).
The AGN space density shows a large variation with redshift (a factor of
several hundred) with a marked peak at z=2-3. Note that only the AGN 
space density derived from a pure density evolution model shows such a large 
variation against cosmic time, while the volume luminosity derived from
a pure luminosity evolution model shows a much smaller variation
(see e.g. Boyle and Terlevich~1997). 
The extrapolated comoving volume density of X--ray selected QSO and Seyfert 
galaxies at z=2-3, albeit the systematic uncertainties discussed above,
approaches that of the low-redshift normal galaxies (see Fig. \ref{MAD}a)
as well as that of the
recently discovered population of Lyman-limit galaxies at high redshifts
(Steidel et al.~1996).
This indicates that a substantial fraction of normal galaxies
may host a central supermassive black hole, which turns into an AGN as soon
as it is fuelled, e.g. by interactions. The amount of interactions is much
larger at higher redshifts than today, which can explain the sharp drop
of activity towards low z.
Dormant remnants of AGN are indeed found in almost every nearby galaxy with a
spheroidal component. A tight relation between black hole mass and bulge mass
has been found, indicating that there is a causal connection between the size
of the central black hole and the number of stars in a galaxy (Faber et al., 
1997).

Recently the history of global star formation in the universe has been 
determined e.g. through observations in the Hubble deep field, but also from 
the chemical evolution of Ly$_\alpha$ clouds.
Fig. \ref{MAD}b gives the starforming rate in arbitrary units as a 
function of redshift (Madau 1996, Connolly et al.~1997), 
corrected for dust obscuration following Pettini et
al. (1997). This function shows
only a moderate variation with cosmic time, with a possible peak around a
redshift of unity and a steep decline towards lower redshifts, the shape
 of which is strikingly similar to the 
low-redshift behaviour of the AGN volume density.

The pronounced peak of the AGN population appears to be at higher redshift 
than the maximum of the star formation,
indicating, that the bulk of black holes which nowadays form the
center of normal galaxies has been in place before the bulk of the stars in
these galaxies have formed. In this picture the X--ray background is just
the total emission of all black hole accretion processes, integrated over
cosmic space and time. Following the argument by Soltan (1982), the
total mass accreted onto black holes integrated over cosmic time can be
estimated from the AGN number counts and thus from the X--ray background 
radiation. This corresponds to an average
mass in dormant black holes of $10^7 - 10^8 M_\odot$ {\it per galaxy} 
(see e.g. Chokshi \&
Turner 1992) and is surprisingly close to the average black hole mass measured
for local galaxies (Faber et al.~1997).

The scenario described here suggests an attractive solution to the X--ray 
background puzzle, where the XRB is just the ``echo'' of mass accretion 
processes onto supermassive black holes in the centers of nearly 
all galaxies.
However, only deeper survey observations with the next generation
of sensitive X--ray observatories with good angular resolution in the
harder X--ray band (i.e. AXAF and XMM), where the bulk of the 
energy density of the XRB resides, and optical follow-up spectroscopy from
8-10m telescopes will be able to unambiguously confirm the AGN XRB model 
and finally solve this long-standing mystery.

\acknowledgements

{\small I acknowledge the grant 50 OR 9403 5 by the Deutsches Zentrum f\"ur
Luft- und Raumfahrt (DLR, former DARA). I also warmly thank my collaborators 
in the ROSAT Deep Survey and ROSAT Bright Survey projects, Thomas Boller,
Jens-Uwe Fischer, Riccardo Giacconi, Ingo Lehmann, Takamitsu Miyaji, Maarten 
Schmidt, Axel Schwope, Joachim Tr\"umper, Wolfgang Voges and Gianni Zamorani 
for fruitful discussions and the permission
to discuss data in advance of publication. In particular I am grateful to
M.~Schmidt and T.~Miyaji for help with the luminosity function.}

\refer
\aba
\rf{Boyle B.J., Shanks T., Georgantopoulos I., Stewart G.C., Griffiths R.E., 
    1994, MNRAS 260, 49}
\rf{Boyle B.J., McMahon R.G., Wilkes, B.J., Elvis M., 1995, MNRAS 272, 462}
\rf{Boyle B.J., Terlevich R.J., 1998, MNRAS submitted (astro-ph/9710134)}
\rf{Branduardi-Raymont G., Mason K.O., Warwick R.S., et al. 1994, MNRAS 270, 
    947}
\rf{Chokshi A., Turner E.L., 1992, MNRAS 259, 421}
\rf{Comastri A., Setti G., Zamorani G., Hasinger G., A\&A 296, 1}
\rf{Connolly A.J., Szalay A.S., Dickinson M., Subbarao M.U., Brunner R.J., 
    1997, ApJ 486, L11}
\rf{de Ruiter H., Zamorani G., Parma P., et al., 1997, A\&A 319, 7}
\rf{Faber S.M., Tremaine S., Ahjar E.A. et al., 1997, AJ in press 
  (astro-ph/9610055)}
\rf{Georgantopoulos I., Stewart G.C., Shanks T., Boyle B.J., Griffiths R.E.,
    1996, MNRAS 280, 276}
\rf{Gioia I.M., Maccacaro T., Schild R.E., et al., 1990, ApJ Suppl. 72, 567}
\rf{Griffiths R.E., Della Ceca R., Georgantopoulos I., et al.,
    1996, MNRAS 281, 71}
\rf{Hasinger G., Burg R., Giacconi R., et al.,
    1993, A\&A 275, 1}
\rf{Hasinger G., Burg R., Giacconi R., et al.,
    1998a, A\&A, in press (astro-ph/9709142)}
\rf{Hasinger G. et al., 1998b, AN, in prep.}
\rf{Inoue H., Kii T., Ogasaka Y., Takahashi T., Ueda Y., 1996, MPE report 263, 
    323}
\rf{Jones L.R., McHardy I.M., Merrifield M.R. et al., 1997, MNRAS 285, 547}
\rf{Madau P., 1996, astro-ph 9612157}
\rf{Mason, K. et al., 1996, priv. comm}
\rf{McHardy I.M., Jones L.R., Merrifield M.R. et al., 1997, MNRAS, in press, 
    (astro-ph/9703163)}
\rf{Page M.J., Carrera F.J., Hasinger G., et al., 1996, MNRAS 281, 579 }
\rf{Pei Y.C., Fall S.M., 1995, ApJ 454, 69}
\rf{Pettini M., Steidel C.C., Adelberger K.L., et al.,
    1997, astro-ph 9708117}
\rf{Rees M., 1989, Reviews of Modern Astronomy 2, 1}
\rf{Roche N., Griffiths R.E., Della Ceca R., et al.,
    1995, MNRAS 273, 15}
\rf{Schmidt K.-H., Boller T., Voges W., MPE report 263, 395}
\rf{Schmidt M., Schneider D.P., Gunn J.E., 1995, AJ 110, 68}
\rf{Schmidt M., Hasinger G., Gunn J., et al. 1998, A\&A, in press 
    (astro-ph/9709144)}
\rf{Schneider D., et al., 1998, AJ submitted}
\rf{Shanks T., Gorgantopoulos I., Stewart G.C., et al.,
    1991, Nature 353, 315}
\rf{Soltan A., 1982, MNRAS 200, 115}
\rf{Steidel C.C., Giavalisco M., Pettini M., Dickinson M., Adelberger K.L., 
    1996, AJ 112, 352}
\rf{Voges W., Aschenbach B., Boller T., et al., 1995, IAU Circ \# 6102,
    \hbox{www.rosat.mpe.mpg.de/survey/rass-bsc}}
\abe

\addresses
\rf{G\"unther Hasinger,
Astrophysikalisches Institut Potsdam,
An der Sternwarte 16,
D-14482 Potsdam,
Germany,
e-mail: GHasinger@aip.de}
END

\end{document}